# Superconductivity above 100 K in PH$_3$ at high pressures

A. P. Drozdov, M. I. Eremets and I. A. Troyan

*Max-Planck Institut fur Chemie, Hahn-Meitner Weg 1,55128, Mainz, Germany*

Following the recent discovery of very high temperature conventional superconductivity in sulfur hydride (critical temperature T$_c$ of 203 K, Ref[1]) we searched for superconductivity in other hydrides and found that a covalent hydride phosphine (PH$_3$) also exhibits a high T$_c$ >100 K at pressure P > 200 GPa as determined from four-probe electrical measurements.

.

The recent finding of superconductivity in H$_x$S system at 200 K under pressure[1] experimentally supports a statement of BCS theory in Eliashberg extension[2] that there is no apparent limit for the critical temperature (T$_c$) of superconductivity[3] and it stimulates the search for even higher transition temperatures. Hydrogen rich materials (and metallic hydrogen) are the most attractive candidates for high and possibly room temperature superconductivity as proposed by N. Ascroft[4,5] because hydrogen atoms provide high frequencies in the phonon spectrum and a strong electron phonon coupling. A lot of theoretical simulations yielding very high superconducting temperatures were made after that predictions (see recent references[6-9]). Experimentally superconductivity was found in SiH$_4$ with T$_c$ = 17 K (Ref [10]), in BaReH$_9$ (Ref [11]) with T$_c$ ~ 7 K, and finally at 200 K in sulfur hydride[1]. Other hydrogen containing materials such as AlH$_3$ (Ref [12]), B$_{10}$H$_{14}$ (Ref [13]) however did not reveal superconductivity. Thus there is still the question whether the high temperature superconductivity in sulfur hydride system is a unique phenomenon among hydrogen containing materials.

We selected PH$_3$ for the present study because this material with its covalent bonding is similar to H$_2$S. Few experimental studies were performed on this molecular crystal such as X-ray diffraction[14], infrared absorption and Raman scattering[15]. To our knowledge there are no studies under pressure both in experiment and theory.

Phosphine gas (99,999% from Air Liquide) was liquiditied in a diamond anvil cell (DAC) at 170 K (Fig. 1a) using the same technique as in Ref. [1]. Presence of the sample was controlled visually and by measuring Raman spectra showing the characteristic peak of PH$_3$ at ~2300 cm$^{-1}$ (Ref. [15]). After clamping the sample it was pressurized at low temperatures of T < 200 K to avoid possible decomposition. Electrical measurements were performed with four electrodes in the van der Pauw geometry with the same technique as in Ref. [1].

Under pressure the P-H stretching mode of phosphine at 2300 cm$^{-1}$ shifts to higher frequencies. The Raman intensity decreases with pressure and the signal vanishes, with the sample darkening at pressure >25 GPa. The sample becomes opaque and starts to conduct at ~30 GPa (measured at 180 K), and weak visible reflection appears at 35 GPa. We did find indication for a semiconducting behavior: photoconductivity was not observed at 40 GPa, indicating that the sample is already metallic. We did not measure the temperature dependence of resistance at this pressure to check, if the sample is metallic and possibly superconducting. Temperature scans were measured at pressures P>80 GPa (Fig. 1b-d).

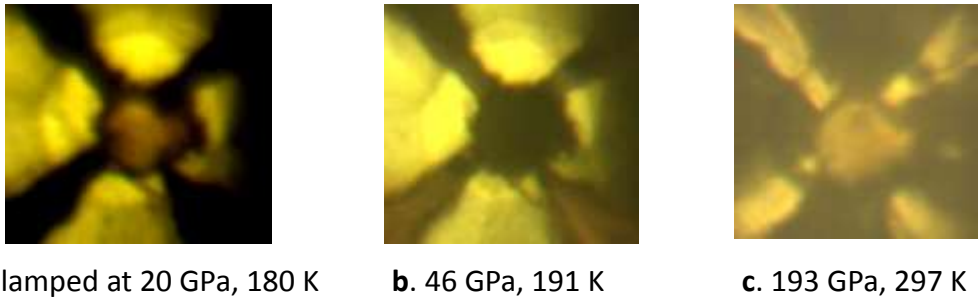

**a**. Clamped at 20 GPa, 180 K     **b**. 46 GPa, 191 K     **c**. 193 GPa, 297 K

Fig. 1. Photographs of the sample at different pressures and temperatures viewed through the diamond anvils. **a** The clamped sample. It is seen as brown disk on the yellow circle (diamond culet), four electrodes touch it. The photograph was taken in transmitted light. **b**. At 46 GPa the sample is opaque and well conducts electrical current (resistance is about 500 Ohm). **c**. The sample seen in reflected light.

At cooling at 83 GPa we observed a signature of a superconducting transition: the resistance abruptly dropped to "zero" at the critical temperature ($T_c$) of 30 K, Fig. 1b. Resistivity of the sample below the transition temperature is about $10^{-9}$ Ohm m which is the limit of our measurement device (Fig 1e). Next, pressure was increased at T <200 K and then the sample was cooled down. The critical temperature $T_c$ increased with pressure up to 103 K at 207 GPa (Fig. 2b). This pressure run was reproduced twice at similar temperature conditions of loading (Fig. 1cd). The dependence of the critical temperature $T_c$ on the pressure is plotted in Fig. 1e.

These first experimental data clearly evidence another (besides hydrogen sulfide) hydride with very high $T_c$. It is likely conventional superconductor. Obviously the experiments should be extended to studies of the isotope effect in $PD_3$ and of magnetic susceptibility.

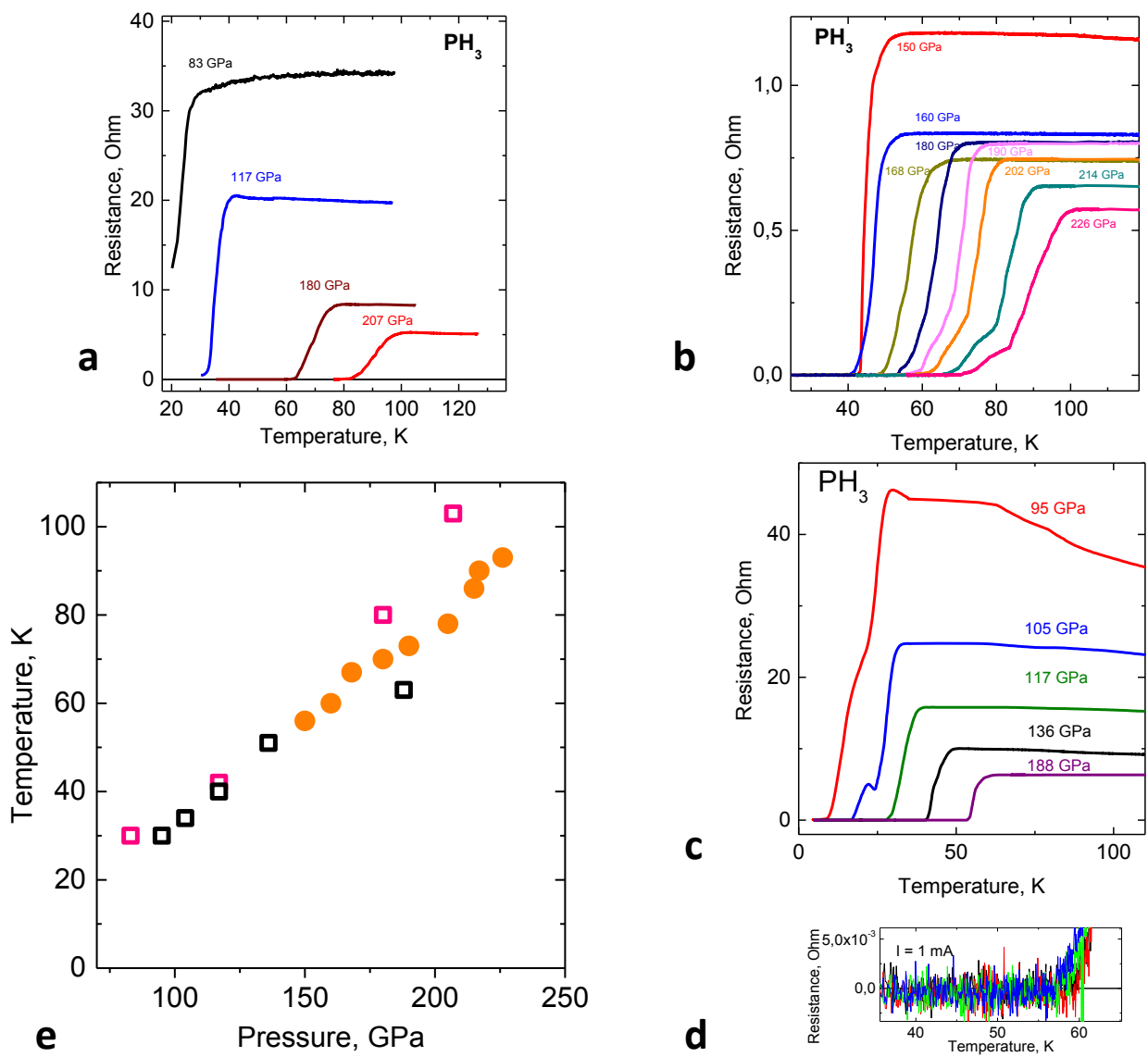

Fig. 2. Electrical measurements of phosphine at high pressures. **a-c** Superconducting edges as seen in different runs in the temperature dependence of resistance at different pressures. The curves were obtained at slow (hours) warming of the sample to ensure correct temperature measurements. In the superconducting state resistance dropped to zero as scaled in **d**. **e**. Dependence of the critical temperature $T_C$ on pressure. The points were taken from plots **a-c.** The critical temperature $T_c$ is defined here as the sharp kink in the transition to normal metallic behavior.